%% file: main.tex
\documentclass[conference]{IEEEtran}

\input{src/preamble.tex}
\begin{document}


\title{Trim My View: An LLM-Based Code Query System for Module Retrieval in Robotic Firmware}
\author{
  \IEEEauthorblockN{
    Sima Arasteh\IEEEauthorrefmark{1},
    Pegah Jandaghi\IEEEauthorrefmark{2},
    Nicolaas Weideman\IEEEauthorrefmark{2},
    Dennis Perepech\IEEEauthorrefmark{1}, \\
    Mukund Raghothaman\IEEEauthorrefmark{1},
    Christophe Hauser\IEEEauthorrefmark{3} and
    Luis Garcia\IEEEauthorrefmark{4}}
  \IEEEauthorblockA{
    \IEEEauthorrefmark{1}University of Southern California, \\
    \url{{arasteh, perepech, raghotha}@usc.edu}}
  \IEEEauthorblockA{
    \IEEEauthorrefmark{2}University of Southern California / Information Sciences Institute, \\
    \url{jandaghi@usc.edu}, \url{nhweideman@gmail.com}}
  \IEEEauthorblockA{
    \IEEEauthorrefmark{3}Dartmouth College, \\
    \url{christophe.hauser@dartmouth.edu}}
  \IEEEauthorblockA{
    \IEEEauthorrefmark{4}University of Utah Kahlert School of Computing, \\
    \url{la.garcia@utah.edu}}
}

\IEEEoverridecommandlockouts
\makeatletter\def\@IEEEpubidpullup{6.5\baselineskip}\makeatother
\IEEEpubid{\parbox{\columnwidth}{
  Workshop on Binary Analysis Research (BAR) 2025 \\
  28 February 2025, San Diego, CA, USA \\
  ISBN 979-8-9919276-4-2 \\
  https://dx.doi.org/10.14722/bar.2025.23xxx \\
  www.ndss-symposium.org
}
\hspace{\columnsep}\makebox[\columnwidth]{}}

\maketitle


\input{src/abstract.tex}   \mclearpage
\input{src/intro.tex}      \mclearpage
\input{src/related.tex}    \mclearpage
\input{src/chatcps.tex}    \mclearpage
\input{src/setup.tex}      \mclearpage
\input{src/result.tex}     \mclearpage
\input{src/limitation.tex} \mclearpage
\input{src/future.tex}     \mclearpage
\input{src/conclusion.tex} \mclearpage

\bibliographystyle{IEEEtran}
\bibliography{references}

\end{document}

%% file: src/preamble.tex
\usepackage{amsmath}
\usepackage{array}
\usepackage{calc}
\usepackage{comment}
\usepackage{enumitem}
\usepackage{graphicx}
\usepackage[hidelinks]{hyperref}
\usepackage{multirow}
\usepackage{url}
\usepackage{xcolor}

\newcommand{\mclearpage}{}

%% file: src/abstract.tex
\begin{abstract}
The software compilation process has a tendency to obscure the original design of the system and
makes it difficult both to identify individual components and discern their purpose simply by
examining the resulting binary code. Although decompilation techniques attempt to recover
higher-level source code from the machine code in question, they are not fully able to restore the
semantics of the original functions. Furthermore, binaries are often stripped of metadata, and this
makes it challenging to reverse engineer complex binary software.

In this paper we show how a combination of binary decomposition techniques, decompilation passes,
and LLM-powered function summarization can be used to build an economical engine to identify
modules in stripped binaries and associate them with high-level natural language descriptions. We
instantiated this technique with three underlying open-source LLMs---CodeQwen, DeepSeek-Coder and
CodeStral---and measured its effectiveness in identifying modules in robotics firmware. This
experimental evaluation involved 467~modules from four devices from the ArduPilot software suite,
and showed that CodeStral, the best-performing backend LLM, achieves an average F1-score of 0.68
with an online running time of just a handful of seconds.
\end{abstract}

%% file: src/intro.tex
\section{Introduction}
\label{intro}


Knowledge of the high-level organization of software---its functions, files and modules---and how
its functionality is divided among these units is crucial to productive programming,
reverse engineering, and other software engineering practices. While at least some of this
information is evident while examining the \emph{source code}, it is much more challenging to
recover when working with \emph{binaries}.

Crucially, the compilation process tends to obscure the organizational units of software, including
information about its modules, comments, data types, and identifier names (when working with
stripped binaries). Although this is sometimes desirable (for example, to protect intellectual
property), it severely hinders the understandability of the final binary.
Furthermore, despite impressive advances in decompilation techniques~\cite{liu2020far}, they are unable to
completely recover semantic details of the original source code~\cite{cao2024evaluating}.


Recent breakthroughs in the development of large language models (LLMs) have prompted researchers to
investigate their application in various problems related to the semantic understanding of (binary)
code: i.e., binary code summarization~\cite{jin2023binary, song2024bin2summary}, source-to-binary
matching~\cite{jiang2024binaryai}, and function identification~\cite{chen2024foc}. Several studies
have also evaluated their effectiveness for function summarization tasks~\cite{jin2023binary,
song2024bin2summary, al2023extending, mastropaolo2024evaluating, ahmed2022few}.


Despite their impressive accuracy, we argue that function summarization is insufficient for a number
of problems related to understanding the high-level structure of programs. For example, engineers
may wish to identify portions of the codebase responsible for implementing communication protocols
such as the serial peripheral interface (SPI), or for managing the universal asynchronous
receiver/transmitter (UART). These functionalities may be deeply embedded in the code and/or
divided among multiple functions. This limits the effectiveness of function summarization tools for
many reverse engineering activities, which instead require identifying and navigating between larger
units of the programs, such as its modules.

Another challenge that arises while using LLMs is economic in nature: Well-known commercial LLMs
such as GPT-X~\cite{achiam2023gpt}, Copilot,\footnote{\url{https://copilot.github.com/}} and
Gemini~\cite{anil2023gemini} charge money when used on a large scale. As an example, the authors of
BinSum~\cite{jin2023binary} report incurring an expense of \$11,000 when evaluating GPT-4. One could
use open-source LLMs instead~\cite{bai2023qwen, zhu2024deepseek}. However, GPUs with large amounts
of memory are prohibitively expensive, and this would restrict us to use mid-size language models.
This raises an additional worry that these smaller locally-run models would be less accurate or
effective at discharging their tasks.


In this paper, we describe a system---called ChatCPS---in which the user requests parts of the code
related to some high-level functionality of interest. The code takes the form of stripped binaries.
In response, the system automatically identifies modules/groups of related functions related to
the desired functionality and presents them to the user for examination. As indicated by the name
(\emph{$\cdots$-CPS}), our primary focus is on reverse engineering firmware for robot controllers.

\begin{figure*}
\centering
\includegraphics[width=.7\textwidth]{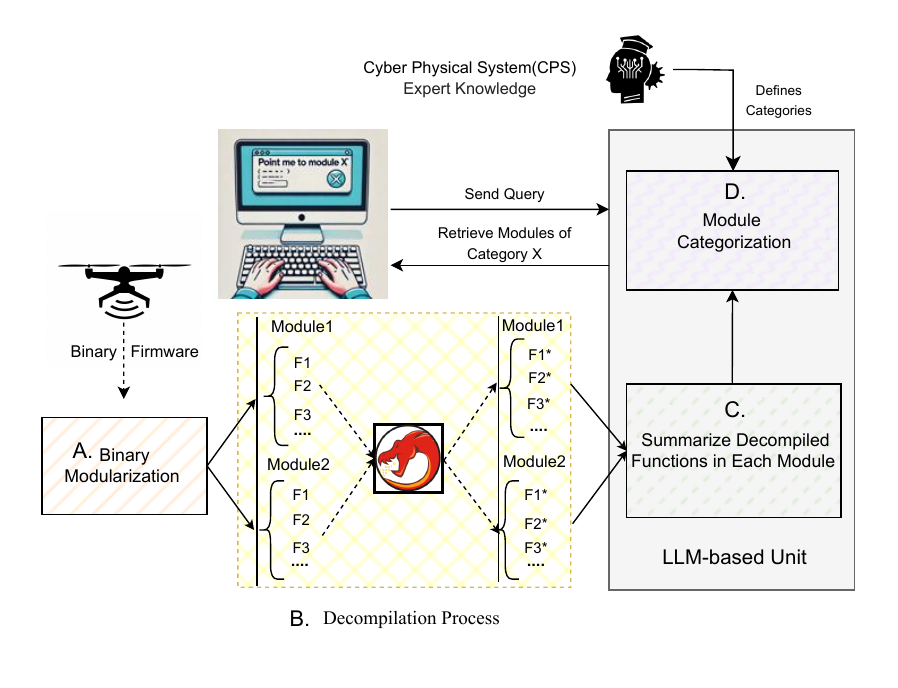}
\caption{The Workflow of the ChatCPS system. A) ChatCPS decomposes the binary firmware into its
  modules. B) Shows the decompilation process. $F_n$ indicates binary functions, while $F_n^*$
  indicates the corresponding decompiled functions. C) ChatCPS summarizes each function in a module
  using three open-source LLMs (CodeQwen, CodeStral, DeepSeek-Coder) and D) categorizes modules
  based on function descriptions.}
\label{fig:chatcps-workflow}
\end{figure*}

Our system relies on a combination of binary decomposition techniques, function decompilation, and
multiple levels of LLM application. In the first LLM pass, we obtain summaries of individual
functions with the binary. In parallel, we apply BCD~\cite{karande2018bcd}, an algorithm for
decomposing binary code into components/modules. (Because the source code of the original BCD
implementation was unavailable, and because we were working with ARM rather than x86-64 binaries, we
had to reimplement this algorithm). We then group together summaries for related functions, and use
a second pass with the LLM to obtain a categorization of the module as a whole. We summarize this
workflow in Figure~\ref{fig:chatcps-workflow}. We focus on four categories in which CPS experts
frequently express interest: data transfer, navigation, controllers, and safety checks. We describe
these categories in more detail in Section~\ref{sec:chatcps:module_ret}.

In our experience, this two-pass approach---first summarize individual decompiled functions, and
then summarize groups of related function summaries---improves classification accuracy when compared
to a hypothetical single-pass procedure (which would presumably attempt to directly summarize entire
modules that were previously identified by BCD). This improvement in accuracy allows us to use less
expensive (i.e., smaller and open-source) LLMs, rather than expensive commercial offerings.

We instantiated our system with three mid-sized LLMs drawn from the BigCodeBench leaderboard~%
\cite{bigcode2023leaderboard, zhuo2024bigcodebench}: CodeQwen~\cite{hui2024qwen2}, DeepSeek-Coder~%
\cite{zhu2024deepseek} and CodeStral~\cite{chaplot2023albert}. Our experimental evaluation with
467~modules from the ArduPilot software suite~\cite{ardupilot_dataset} indicates that CodeStral
provides the most accurate high-level categorization, with an average F1-score of 0.68.
As a secondary measurement, we also separately calculated the accuracy of the first LLM pass, i.e.,
we measured the accuracy of the generated function summaries. This measurement required us to build
a source code parser that normalized function bodies by removing comments, variable and function
names. We describe our experimental setup and our results in Sections~\ref{sec:exprimental_setup}
and~\ref{sec:exper_result} respectively.

Our hope is that ChatCPS would streamline the reverse engineering process and enable users to
efficiently identify and distinguish parts of the codebase related to different functionalities. We
also hope that module categorization would enhance the user's understanding of the binary's
underlying structure. Finally, we expect that this top-level categorization would form a basis for
more fine-grained categorization of the functions within each module. This would help to distinguish
different kinds of algorithms, including Kalman filters~\cite{kalman1960new} from PID controllers~%
\cite{visioli2006practical}, both of which may be found in modules related to navigation and
control. At an even lower level, after groups of functions related to individual algorithms have
been identified, users could run more specific queries.

To our knowledge, this is the first such code query-based system that utilizes LLMs to retrieve
specific functionalities within stripped binaries. We summarize our contributions:
\begin{enumerate}
\item We have developed a query system that defines and retrieves module categories using
  LLM-generated summaries.
\item We implement a new variant of the BCD algorithm adjusted for our needs and evaluate its
  effectiveness in decomposing robotic firmware into modular components within cyber-physical
  systems.
\item We evaluate the effectiveness of three open-source LLMs (DeepSeek-Coder, CodeQwen and
  CodeStral), in summarizing decompiled functions within modules for stripped firmware.
\item We establish the ground truth categories for 467 modules across four devices in the
  ArduPilot~\cite{ardupilot_dataset} dataset, including QuadCopter, HeliCopter, Rover, and
  Submarine.
\end{enumerate}
The ChatCPS implementation may be downloaded from \url{https://github.com/SimaArasteh/chatcps}.

%% file: src/related.tex
\section{Related Work}
\label{sec:related}

Code semantics analysis has been widely studied by researchers. Traditionally, researchers have relied on signature-based methods to detect specific, well-known functions, such as mathematical operations. For example, advanced binary analysis tools like IDA Pro~\cite{eagleida}, Ghidra~\cite{ghidra}, and angr~\cite{shoshitaishvili2016sok} are designed to recognize the usage of common libraries within binaries, such as standard math functions. More sophisticated techniques employ semantic pattern matching to identify known functions. Studies such as those by Kim et al.~\cite{kim2020revisiting} and Xu et al.~\cite{xu2017cryptographic} have focused on extracting and recognizing semantic patterns associated with specific mathematical functions, including cryptographic operations.


However, these traditional approaches are generally constrained to a predefined set of patterns and are incapable of detecting unknown functions.

To address these limitations, recent research has begun leveraging LLMs. LLMs have demonstrated remarkable effectiveness across a variety of tasks in the area of software engineering, including function identification and code summarization. 
For example, BinaryAI~\cite{jiang2024binaryai} leverages LLMs to correlate binary code snippets with their original source code, aiding in tasks like detecting reused third-party libraries.
Similarly, FoC~\cite{chen2024foc} focuses on identifying known cryptographic functions within stripped binaries via LLMs.

Moreover, LLMs can extract code semantics and software functionality. Before the widespread adoption of LLMs, researchers used smaller transformer models such as CodeT5~\cite{wang2021codet5}, BinT5~\cite{al2023extending} and CodeBERT~\cite{feng2020codebert} for code summarization tasks to enhance comprehension of both binary and source code~\cite{huanzhen2023improve, mastropaolo2024evaluating, arakelyan2023exploring, mondal2023understanding, al2023extending}. Smaller transformer models like CodeT5, BinT5, and CodeBERT may offer less detailed code summaries due to their limited parameters, which can restrict their ability to handle diverse coding languages and scenarios effectively. Larger LLMs, with their extensive training and advanced architectures, provide more accurate and comprehensive summaries. As a result, researchers have begun to explore the capabilities of larger LLMs in understanding the code semantics. For example,~\cite{jin2023binary, ahmed2022few, sun2024source, haldar2024analyzing, lomshakov2024proconsul} and~\cite{arakelyan2023exploring}  conducts evaluations of different LLMs such as ChatGPT,  Llama2~\cite{rejithkumar2024towards}, and Code-Llama~\cite{roziere2023code} for the task of code summarization.

While code semantic summarization is useful for understanding individual components of a software system, it does not offer a comprehensive view of the software's overall architecture and interactions. In reverse engineering, especially with large and complex software, analyzing every function can be overwhelming and unnecessary. Instead, focusing on specific functionalities that are crucial can be more effective. Developing a system that selectively highlights these key functionalities would greatly enhance the efficiency of the analysis.

\begin{figure}[ht]
\centering
\includegraphics[width=\columnwidth]{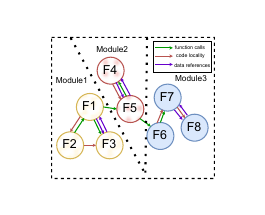}
\caption{Illustration of how the BCD algorithm uses information about the call graph, data
  references, and code locality to decompose the binary into groups of functions. $F_n$ indicates
  functions in the binary. The image shows a final weighted graph generated by the BCD algorithm.
  Green edges show the function call relationship between functions. Red edges show the code
  locality, and the purple edges indicate the data references.}
\label{fig:decomposition}
\end{figure}

%% file: src/chatcps.tex
\section{The Design of ChatCPS}
\label{sec:chatcps}

Reverse engineering complex software without access to the source code is a challenging and labor-intensive task. To facilitate this process, we introduce ChatCPS, a code-based query system designed specifically for cyber-physical systems. This system aims to assist reverse engineers in focusing on critical software functionalities. ChatCPS consists of a three-layer pipeline. First, we decompose binary firmware into its software components called modules. Then, we summarize the decompiled version of functions in the modules using LLMs. Finally, we categorize modules based on the function descriptions by leveraging LLMs. Figure~\ref{fig:chatcps-workflow} shows the overall workflow of ChatCPS.

\begin{figure*}[ht]
  \centering
  \includegraphics[width=\textwidth,height=\textheight,keepaspectratio]{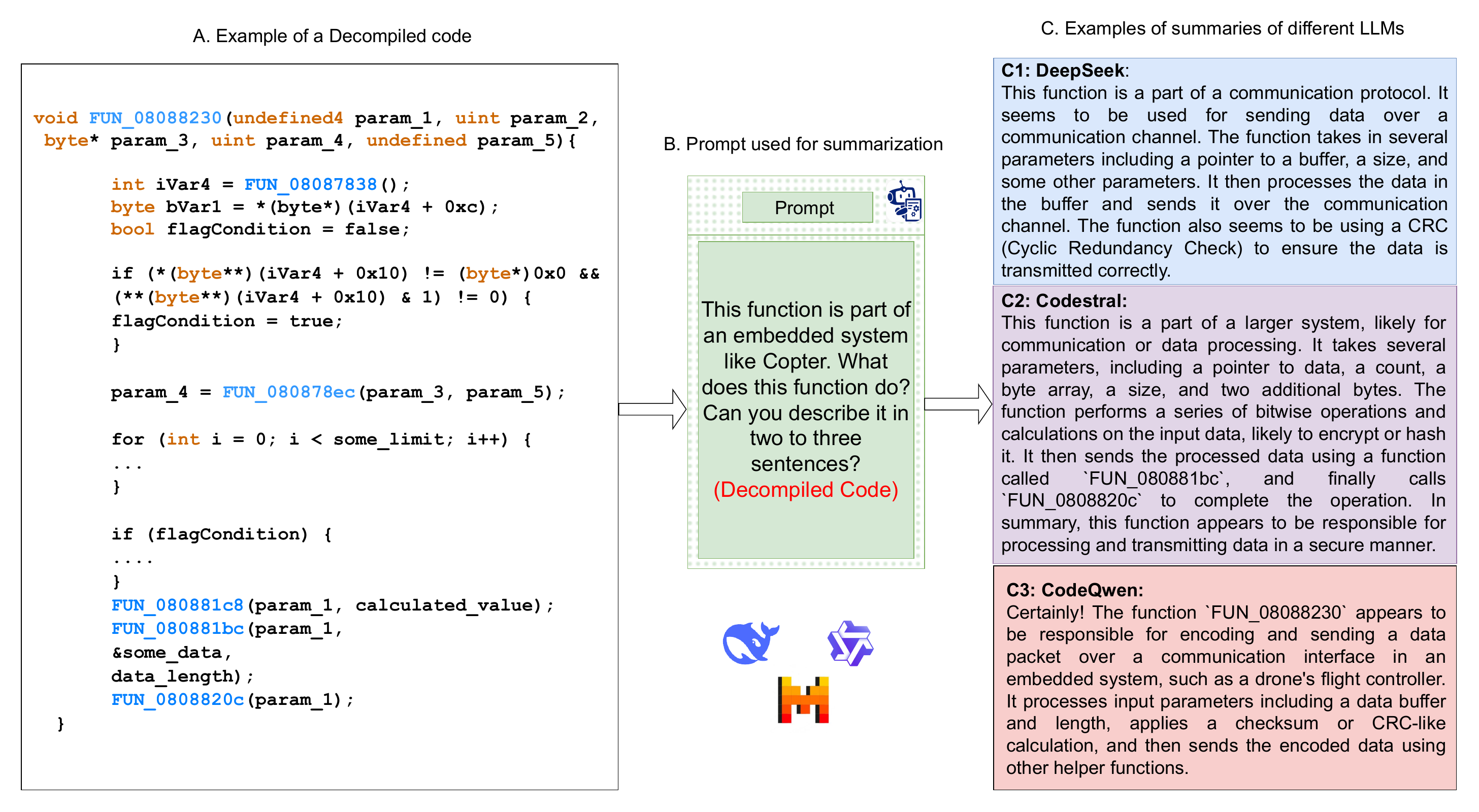}
 \caption{The process of summarizing decompiled functions using LLMs. Section A-- presents a decompiled example of the function \textit{\_mav\_finalize\_message\_chan\_send} from the ArduPilot project, extracted from the stripped binary of a copter device compiled for the fmuv3 board. Section B-- depicts the prompt used to summarize decompiled functions. Section C-- displays the function descriptions of \textit{\_mav\_finalize\_message\_chan\_send}, as returned by three distinct LLMs.}
  \label{fig:summ}
\end{figure*}

\begin{figure*}[ht]
  \centering
  \includegraphics[width=\textwidth,height=\textheight,keepaspectratio]{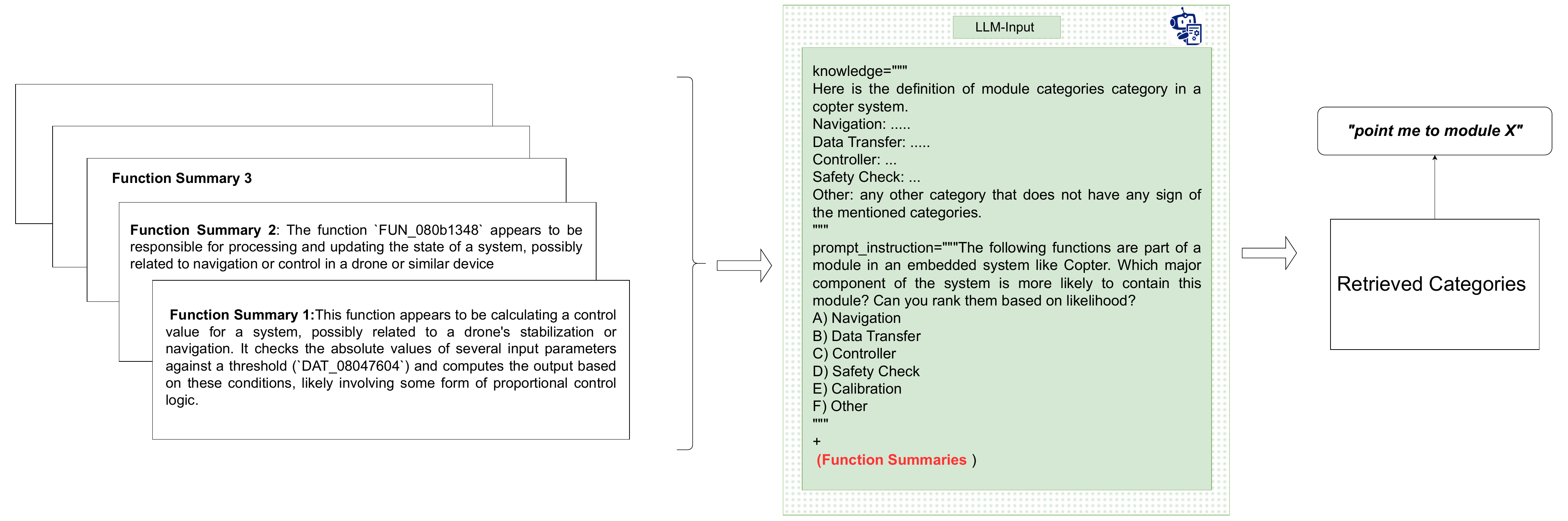}
 \caption{Module Retrieval Process. We provide function summaries and module categories as an input prompt to an LLM. Then, we instruct the LLM to decide based on the function descriptions and ranked the categories. }
  \label{fig:category}
\end{figure*}

\subsection{Binary Modularization}
\label{sec:chatcps:modularization}
We start by decomposing binary firmware into its constituent modules using a modified version of the BCD method as described by Karande et al.~\cite{karande2018bcd}. While the original source code was unavailable and required us to re-implement the technique, the modifications were made specifically to address the incompatibility of the original BCD method with ARM binaries. In this section, we outline the original workflow of the BCD method and detail the changes we introduced to adapt it for ARM binaries.

To decompose a binary into its modules, the BCD method constructs a directed graph from the binary program, where nodes represent functions and edges capture three types of relationships between functions: code locality, data references, and function calls. It then applies the Newman algorithm~\cite{newman2004fast} to cluster the graph into smaller components, referred to as modules.

BCD relies on three facts to decompose a binary program.

\subsubsection{\textbf{Code Locality}} 
\label{sec:chatcps:modularization:codelocality}
Functions within a module typically share similar functionalities and are physically located close to each other. This proximity is maintained by the compiler in the binary program, where functions are organized sequentially from lower to higher memory addresses. BCD utilizes this characteristic to construct a directed graph known as \textit{SG} (for \emph{sequence graph}), which illustrates these relationships. In this graph, an edge extends from a function at a lower memory address to a function at a higher address, effectively mapping the sequential organization of functions within the binary.

\subsubsection{\textbf{Data References}}
\label{sec:chatcps:modularization:datareferences}
Functions within a module often access and share the same data. BCD examines this access for global and static variables, as well as constant string literals located in the \textit{.data}, \textit{.bss}, and \textit{.rodata} sections. In this process, BCD constructs a graph known as the Data Reference Graph (\textit{DRG}). In this graph, an edge is drawn between two functions if they access at least one common variable. Consequently, this shared access results in a mutual edge being established between these functions within the graph.

\subsubsection{\textbf{Function Calls}}
\label{sec:chatcps:modularization:functioncalls}

The principle underlying BCD is the observation that functions within a module tend to call each other more frequently than they do functions outside the module. To leverage this insight, BCD constructs a graph called the Call Graph \textit{CG}, wherein it assigns a weight to each edge based on the number of calls between functions. Additionally, BCD synthesizes a comprehensive weighted graph by integrating the CG with two other graphs:  (\textit{SG}) and (\textit{DRG}).

In this composite graph, the weight of each edge is determined by a linear combination of its weights from SG, DRG, and CG. During the final stage of the process, BCD employs an algorithm known as the Newman algorithm to cluster the graph into smaller graphs refer as a module.
Based on this algorithm, each node is considered an independent cluster. Subsequently, nodes are merged based on the optimal connectivity indicated by their edges, ultimately forming the final clusters.
For a detailed explanation of this methodology, we refer the reader to the original paper~\cite{karande2018bcd}. Figure~\ref{fig:decomposition} shows the workflow of the BCD algorithm.

\subsection{Function Summarization in Binary Modules}
\label{sec:chatcps:func_sum}

In this step, we leverage open-source LLMs to summarize decompiled functions in each module. Within our codebase, some functions are relatively small, primarily serving as interfaces that call numerous other functions. These smaller functions typically reveal less about the program's semantic structure. To focus on more substantively informative functions, we exclude functions with a line count below a specified threshold. This approach ensures that our analysis concentrates on functions that provide significant insights into the software's architecture. Figure~\ref{fig:summ} shows the prompt and the process that we used for summarizing functions. In this Figure, you can observe summaries generated by three open-source  LLMs(Codeqwen, Codestral, and DeepSeek-Coder) for function \textit{mav\_finalize\_message\_chan\_send} within a module of a helicopter device. 

\subsection{Module Retrieval}
\label{sec:chatcps:module_ret}
In this step, our goal is to identify modules with specific functionalities within cyber-physical systems. Given their wide range of applications and unique capabilities, these systems serve as our primary focus, though our methodology is applicable to various types of applications. We define five distinct module categories based on common functionalities across cyber-physical systems (e.g., robots): 
1) Data transfer (crucial for real-time communication between vehicle sensors and systems), 2) navigation (essential for route optimization and precise positioning), 3) control (manages operational capabilities such as steering and braking), and 4) safety checks (monitors systems and environment to prevent accidents). We also consider the modules that do not align with these defined categories and label them with category \texttt{other}. Notably, these categories are not mutually exclusive, as some modules belong to multiple categories, adding complexity to the module retrieval process.

To identify the module category using LLMs, we construct a prompt for each module. The prompt contains three sections to provide the LLM with sufficient information to accurately infer the category of the module:\\
\textbf{Category Definition/Knowledge}: This section includes the precise definition of module categories, which are curated by an expert. We include this section to ensure that LLM uses the intended semantics of categories rather than relying on their semantic prior. \\
\textbf{Functions Summaries}: In this section, we include the summaries of the functions contained in the module, which are generated in the previous stage of the pipeline.\\
\textbf{Prompt Instruction}: This section provides the LLM with specific instructions, including the context of the code and a clear description of the task it is expected to perform. Specifically, these instructions guide the LLM to rank the categories based on the likelihood of the module belonging to each. This approach also allows the LLM to retrieve multiple categories when a module aligns with more than one. Figure~\ref{fig:category} shows the prompt and the process we use for this phase.


%% file: src/setup.tex
\section{Experimental Setup}
\label{sec:exprimental_setup}
In this section, we describe the benchmarks and experimental setup, including detailed information on the configurations and methods applied in our experiments. 

\subsection{Benchmark}
\label{sec:exprimental_setup:benchmark}

To evaluate our approach, we use the open-source and well-known autopilot dataset called ArduPilot. This dataset is a comprehensive collection of data and configurations that supports the development and testing of autonomous vehicle software across various platforms. Developers created ArduPilot to boost research and development in unmanned vehicle systems, offering a solid base for simulation, testing, and the practical application of autonomous control algorithms. The dataset includes a wide range of vehicles, such as QuadCopters, HeliCopters, Planes, Rovers, and Submarines. As an open-source software, ArduPilot is widely used to innovate and improve unmanned vehicle technology.

We compile the ArduPilot dataset for four devices—including QuadCopters, HeliCopters, Rovers, and Submarines— specifically for the FMUV3 board. The FMUV3 is a critical flight management unit version that supports various processors and sensors. This versatility enhances its utility across different unmanned vehicle platforms within the ArduPilot framework. Though Ardupilot provides many different boards, we only focus on building the ArduPilot dataset for just one board due to the inherent challenges of establishing ground truth for all boards. Also, Different boards share common functionalities and modules. 
We plan to extend our analysis to additional boards in future work.

\subsection{Binary Modularization Setup}

We decompose binaries into modules using the BCD algorithm~\cite{karande2018bcd}. As the original source code was unavailable, we adapted the approach outlined in the paper with some modifications. Originally, BCD was based on IDA-Pro~\cite{eagleida}, which is not free. Therefore, we opted to implement BCD using angr~\cite{shoshitaishvili2016sok}. Also, the original BCD algorithm was designed for x86 binaries. However, as ArduPilot binaries utilize the ARM architecture, we modify the BCD implementation to align with our dataset. This adaptation specifically addresses the differences in CPU architecture, which affect the identification and processing of shared data references between functions. Due to the distinct ways ARM and x86 architectures handle these references, our implementation requires customization to effectively manage these architectural variations.

\begin{table}
\renewcommand{\arraystretch}{1.3}
\centering
\caption{The effectiveness of the BCD algorithm in decomposing binary firmware into its modules in ArduPilot.}
\label{tab:modularization-effectiveness}
\footnotesize 
\setlength{\tabcolsep}{3pt} 
\begin{tabular}{cccccc} 
\hline
\textbf{Device} & \textbf{\# of Modules} & \textbf{\# of Functions} & \textbf{$\mathbf{P_w}$} & \textbf{$\mathbf{R_w}$} & \textbf{$\mathbf{F1_w}$} \\ 
\hline
QuadCopter & 121 & 8556 & 0.77 & 0.69 & 0.72 \\
HeliCopter & 121 & 7583 & 0.75 & 0.73 & 0.74 \\
Rover & 112 & 7258 & 0.72 & 0.72 & 0.72 \\
Submarine & 113 & 6994 & 0.76 & 0.68 & 0.71 \\
\hline
\end{tabular}
\end{table}

\subsection{Decompilation Process Setup}

We decompile binary functions extracted from stripped binaries in ArduPilot using Ghidra~\cite{ghidra}, a free and open-source reverse engineering tool developed by the National Security Agency (NSA). Ghidra offers a wide range of features for analyzing compiled code across multiple platforms and architectures.

For each module, we select functions that contain at least 15 lines of code to ensure they are large enough to provide meaningful semantic insights. Some functions in a module are limited to calling other internal or external functions, providing limited semantic insight by themselves. As a result, we filter out such functions in our evaluation. Furthermore, we focus solely on the body of the function itself without delving into the bodies of callee functions. This approach is justified for several reasons. Firstly, larger functions typically contain enough content to disclose their semantic purpose. Secondly, a single function might invoke numerous other functions with entirely different semantics. For example, a controller function might call a data-transfer function to send a message. Including the bodies of these callee functions could skew the results of our categorization and potentially mislead our analysis.

\subsection{LLM-based Unit Setup}

To retrieve modules with specific functionalities, we evaluate three open-source LLMs: Deepseek-Coder\cite{zhu2024deepseek}, CodeQwen\cite{hui2024qwen2}, and CodeStral~\cite{chaplot2023albert}. DeepSeek-Coder, CodeQwen, and CodeStral are specialized versions of well-known language models designed for code analysis and development tasks. DeepSeek excels at interpreting both language and code, significantly aiding in the review of complex code and the creation of documentation. CodeQwen, built to improve coding help, provides features such as real-time debugging and code optimization. CodeStral builds on the capabilities of the Mistral model family and specializes in analyzing code in multiple languages. It supports a variety of programming environments and helps ensure that code works well across different languages.

Based on their good performance on the CodeBench leaderboard~\cite{bigcode2023leaderboard}, we selected these LLMs for our study and plan to analyze additional LLMs in future work. To ensure a fair comparison, we chose models with similar parameter counts, specifically Qwen2.5-Coder-32B-Instruct, DeepSeek-Coder-33B-Instruct, and Codestral-22B-v0.1. The context window for these models is 128000, 16000, and 32000 tokens, respectively. Our experiments were conducted on a CentOS-based server equipped with two GPUs, using parallel processing across both GPUs and featuring 48 GB of RAM.
We take advantage of the versions of these open-source LLMs available through HuggingFace\footnote{https://huggingface.co/}. For inference, we use the widely adopted lm-eval-harness library~\cite{eval-harness}. This evaluation framework provides a streamlined and flexible interface for benchmarking LLM performance across diverse tasks, enabling systematic and reproducible evaluations.

 Our implementation of ChatCPS including the modified version of BCD contains approximately 2000 lines of Python. 


\begin{table}
\renewcommand{\arraystretch}{1.5}
\centering
\caption{LLM Function Summarization Evaluation. The numbers indicate the average cosine similarity of the encoded summary of the normal code and the decompiled code of the function generated by the LLMs in the columns.  }
\label{tab:summary_effectiveness}
\begin{tabular}{c|c|c|c} 
\hline
Device & Deepseek-Coder & CodeStral & CodeQwen  \\ 
\hline
QuadCopter &     0.76 $\pm$ 0.16           &    0.62 $\pm$ 0.20       &       0.68 $\pm$ 0.20   \\
HeliCopter   &  0.75 $\pm$ 0.18               &    0.62 $\pm$ 0.20       &      0.68 $\pm$ 0.21     \\
Rover  &        0.75 $\pm 0.17$        &        0.62 $\pm$ 0.21   &      0.67 $\pm$  0.21  \\
Submarine    &          0.76 $\pm$   0.16   &      0.62 $\pm$ 0.21     &      0.68 $\pm$ 0.21     \\
\hline
\end{tabular}
\end{table}

%% file: src/result.tex
\section{Experimental Results}
\label{sec:exper_result}
Our assessment of ChatCPS addresses the following research questions:
\begin{enumerate}[label=\textbf{RQ\arabic*.}, ref=\textbf{RQ\arabic*}, leftmargin=\widthof{RQ4.}+\labelsep]
\item \label{enu:experiments:rq:effectiveness} How effectively can BCD decompose binaries into modules within the autopilot dataset?
\item \label{enu:experiments:rq:sum} How effectively do CodeQwen, Codestral, and DeepSeek-Coder create function summaries that demonstrate their proficiency in understanding code?
\item \label{enu:experiments:rq:module} How proficiently does ChatCPS retrieve module categories?
\item \label{enu:experiments:rq:time} How long does it take for each LLM to complete function summarization?
\end{enumerate}

\subsection{\ref{enu:experiments:rq:effectiveness}: Binary modularization Effectiveness}

We evaluate how well the BCD algorithm can decompose Ardupilot binaries into its corresponding modules. Table~\ref{tab:modularization-effectiveness} presents statistics for four devices in ArduPilot, including the number of modules created in the binaries, the number of functions, and the performance of modularization. We compare the predicted modularization with the ground truth. We then define the
true positives (TP) as the fraction of functions that were correctly placed in the current module,
the false positives (FP) as the fraction of functions that were incorrectly placed in the current
module, and the false negatives (FN) as the fraction of functions that were not included in the
current module but were expected to be.
From here, similar to the original paper, we calculate a weighted version of precision, recall, and  $F_1$ score shown in equations~\ref{eq:weighted_precision},~\ref{eq:weighted_recall} and ~\ref{eq:weighted_f1} respectively.

\begin{equation}
P_w = \frac{\sum_{i=1}^{N_c} P_i \cdot n_i}{N_f}
\label{eq:weighted_precision}
\end{equation}

\begin{equation}
R_w = \frac{\sum_{i=1}^{N_c} R_i \cdot n_i}{N_f}
\label{eq:weighted_recall}
\end{equation}

\begin{equation}
F_{1w} = \frac{\sum_{i=1}^{N_c} F_{1i} \cdot n_i}{N_f}
\label{eq:weighted_f1}
\end{equation}

In these formulas, $P_i$, $R_i$, and $F_{1i}$ are the precision, recall, and $F_1$ score for module $C_i$, $n_i$ is the number of functions in module $C_i$, $N_f$ is the total number of functions in all modules, and $N_c$ is the total number of modules. We use weighted metrics due to the significant variation in the number of functions each module contains. This method assigns the appropriate importance to each module based on its size, ensuring that modules with a larger number of functions have a greater impact on the overall metrics. This approach is vital as it prevents smaller modules from disproportionately influencing the results, which could distort the system's perceived effectiveness. By adjusting scores relative to the number of functions, the evaluation more precisely reflects the system's performance across modules of varying sizes and complexities, providing a balanced and detailed assessment of its binary decomposition capabilities.

As suggested by Table~\ref{tab:modularization-effectiveness}, the modified version of the BCD algorithm achieved acceptable performance in decomposing binaries into modules. It is important to note that modularizing large binary firmware is a challenging task, yet our implementation of the BCD algorithm has attained a satisfactory level of effectiveness. Even though recovering modules from binary code may be imperfect, it does not significantly impact the effectiveness of categorization. For instance, two different modules in ArduPilot, such as \texttt{AC\_circle} and \texttt{AC\_WPNav}, can both fall under the same category, namely navigation. Furthermore, upon manual inspection, we find that this imperfection in modularization does not substantially cause functions from different modules that belong to distinct categories to be grouped into the same module.

\subsection{\ref{enu:experiments:rq:sum}: Function Summarization Effectiveness}

\begin{table*}
\renewcommand{\arraystretch}{1.5}
\centering
\caption{The effectiveness of retrieving module categories by three open-source LLMs in the ChatCPS design. Numbers between parentheses indicate the upper bound of module categorization. These numbers show the result of module categorization obtained from normalized source code.}
\label{tab:categorization_effectiveness}
\scalebox{0.90}{
\begin{tabular}{|c|c|ccc|ccc|ccc|} 
\hline
\multirow{2}{*}{Device} & \multirow{2}{*}{Category} & \multicolumn{3}{c|}{DeepSeek-Coder} & \multicolumn{3}{c|}{CodeStral} & \multicolumn{3}{c|}{Code-Qwen}  \\ 
\cline{3-11}
                        &                           & Precision & Recall & F1 Score & Precision & Recall & F1 Score & Precision & Recall & F1 Score \\ 
\hline
\multirow{4}{*}{QuadCopter} & Data Transfer             &   0.33 (0.52)      &  0.14 (0.42)     & 0.20 (0.46)        &  \textbf{0.78 (0.76)}         & \textbf{0.73 (0.92)}       &  \textbf{0.75 (0.83)}       & 0.61 (0.72)          & 0.55 (0.55)      &  0.58  (0.62)      \\
                        & Navigation                &  0.24 (0.25)        & 0.68 (0.86)       & 0.36 (0.39)         &  \textbf{0.63 (0.73)}        &  \textbf{0.70 (0.80)}      &  \textbf{0.66 (0.76)}      &  0.24 (0.30)        & 0.82 (0.93)      & 0.37 (0.46)        \\
                        & Controller                & 0.93 (0.93)         & 0.65  (0.67)     &  0.77 (0.77)       &  \textbf{0.92 (0.94)}         &  \textbf{0.92 (0.82)}      &  \textbf{0.92 (0.87)}       & 0.87  (0.88)        & 0.44 (0.52)      &  0.59  (0.64)      \\
                        & Safety Check              &  0  (0)       &  0 (0)     & 0 (0)        & \textbf{0.25 (0.5)}         &  \textbf{0.16 (0.5)}     &  \textbf{0.2 (0.5)}       &  0  (0)       & 0 (0)      & 0   (0)      \\ 
\hline
\multirow{4}{*}{HeliCopter}   & Data Transfer             & 0.5 (0.66)         & 0.36 (0.54)       &0.42 (0.6)         & \textbf{0.73 (0.77)}         & \textbf{0.73 (0.73)}      &\textbf{0.73 (0.75)}         &  1.0 (1.0)         & 0.25 (0.33)        &  0.4 (0.5)        \\
                        & Navigation                & 0.05 (0.13)          & 0.5 (1)       & 0.09 (0.23 )        & \textbf{0.82 (0.9)}        & \textbf{0.83 (0.9)}      &  \textbf{0.82 (0.9)}      &   0.37  (0.65)      & 0.75 (0.9)      &  0.5 (0.75)       \\
                        & Controller                &  0.8  (0.88)       &  0.4 (0.53)     &  0.53 (0.66)       &   \textbf{0.92 (0.92)}      & \textbf{0.9 (0.9)}      & \textbf{0.91 (0.91)}      &  0.81  (0.84)       &   0.9 (0.91)    &  0.85 (0.87)       \\
                        & Safety Check              &    0 (0)      & 0 (0)      &  0 (0)        &  \textbf{0.38 (0.5)}        &  \textbf{0.33 (0.56)}      &  \textbf{0.35 (0.53)}       &  0 (0)        &  0 (0)      &  0 (0)       \\ 
\hline
\multirow{4}{*}{Rover}  & Data Transfer             &  0.23 (0.71)       & 0.42 (0.45)    &  0.3 (0.55)      &  \textbf{0.72  (0.83)}       & \textbf{0.81 (0.83)}      &  \textbf{0.76 (0.83) }     &   0.5 (0.55)        &   0.5 (0.71)      &   0.5 (0.62)       \\
                        & Navigation                & 0.21  (0.4)     &  0.8  (1.0)    &  0.33  (0.57)     &  \textbf{0.68 (0.70)}         & \textbf{0.68 (0.73)}      & \textbf{0.68  (0.71)}      &      0.55  (0.6)      &  0.83 (0.75)      &    0.66 (0.66)       \\
                        & Controller                &  0.75 (1.0)     & 0.23  (0.53)    & 0.35 (0.69)       & \textbf{0.78 (0.0.89) }        &  \textbf{0.92 (0.92)}     & \textbf{0.85  (0.91)}      &  0.83 (1.0)          &  0.71 (0.75)      &    0.76 (0.85)       \\
                        & Safety Check              &    0 (0)       & 0 (0)       &    0 (0)      &  \textbf{0.38 (0.44)}         & \textbf{0.50 (0.67)}       &  \textbf{0.43 (0.53)}       &     0 (0)      &  0 (0)       &     0 (0)     \\ 
\hline
\multirow{4}{*}{Submarine} & Data transfer          & 0.38 (0.44)          & 0.45 (0.8)       & 0.41 (0.57)         & \textbf{ 0.77 (0.78)}        &   \textbf{0.83 (0.85) }   &  \textbf{0.80 (0.81) }     & 0.53 (0.75)         & 0.41 (0.69)      & 0.46 (0.71)        \\
                           & Navigation             & 0.17 (0.23)          & 1.0 (1.0)      &  0.29 (0.37)        &  \textbf{0.73 (0.8)}        &  \textbf{0.78 (0.81) }     &  \textbf{ 0.75 (0.80)}     &   0.6 (0.73)       &  0.6 (0.58)     & 0.6  (0.64)       \\
                           & Controller             &   1.0   (0.66)     &  0.28 (0.42)     & 0.44  (0.52)       &  \textbf{0.78  (0.81) }      & \textbf{0.96 (0.96)}      &  \textbf{ 0.86 (0.88)}     &  0.77  (0.9)       &  0.87 (0.52)      &  0.82 (0.66)       \\
                           & Safety Check          &   0 (0)       & 0 (0)      &  0 (0)        & \textbf{0.30 (0.44)}          &  \textbf{0.43 (0.80)}      &   \textbf{0.35 (0.57)}      &   0 (0)        & 0 (0)       & 0 (0)         \\
\hline
\end{tabular}
}
\end{table*}

Module categories are established based on function summaries, underscoring the importance of their reliability. Therefore, we evaluate the function summarization generated by LLMs separately. To evaluate the effectiveness of function summaries, we compare the similarity between decompiled function summaries and the summaries derived from the normalized version of the original source code.
The normalized version of the source code contains the original code, but function names and variables are stripped. Removing function names and variable names before using LLMs to summarize code helps isolate the LLM's ability to understand and summarize code based purely on its structure and logic rather than relying on potentially descriptive names. Function names and variables can give significant hints about what the code does, which may lead LLMs to rely more on these cues rather than analyzing the underlying code patterns. To evaluate an LLM's raw capability to understand code, we minimize external influences like descriptive names. Ideally, LLM creates a similar textual summary for the decompiled code and the normalized code of the same function. 

To obtain the normalized version of the source code, we develop a source code parser specifically designed to extract function bodies directly from the source code. Then, we parse the ArduPilot source code~\cite{ardupilot} to extract function bodies, using the Clang~\cite{clang} tool to identify the function and variable instances. We replace these identifiers with anonymized names. Additionally, to align the source code with the decompiled functions, we extract mappings between function names and their addresses from binaries that contain debug symbols (DWARF). Using these mapped names and addresses, we identify the corresponding function bodies in the source code. Lastly, we remove all comments from the source code to focus exclusively on evaluating the code understanding capabilities. 

To compute the textual summary similarity between the normalized code and decompiled code for each function, we encode the text using an embedding model. The embedding model generates a vector for each of the textual summaries. Then, we compare the similarity of generated embedding vectors via cosine similarity. We use \texttt{NV-Embed-v2} \cite{lee2024nvembedimprovedtechniquestraining} due to its high performance in the leaderboard\footnote{https://huggingface.co/spaces/mteb/leaderboard} of embedding model quality\cite{muennighoff2023mtebmassivetextembedding}.
The results of this experiment are represented in Table~\ref{tab:summary_effectiveness}.

\subsection{\ref{enu:experiments:rq:module}: Module Retrieval Effectiveness}
For the evaluation of module retrieval, we establish the ground truth for 476 modules from four devices in ArduPilot: QuadCopter, HeliCopter, Submarine, and Rover. 
We establish ground truth for all modules within the binaries, which requires approximately 100 human hours. We compare the module categories retrieved by the LLMs using this ground truth.
In personal communication, the ArduPilot developers expressed their belief that module
categorization was fundamentally hard, even for developers with expert insight.

In all cyber-physical systems, certain functionalities are essential and typically implemented within components known as modules. For clarity and structured analysis, we define four common module categories within this domain. Autopilot devices, for instance, include modules dedicated to communication protocols such as MAVLink, which facilitates the exchange of data between the device and ground stations or other Unmanned Aerial Vehicles (UAVs). These communication modules ensure seamless integration and coordination across the network.

Navigation modules play a crucial role in determining and maintaining the position and orientation of the vehicle, integrating sensors and GPS data to guide its path accurately. Additionally, control modules are fundamental in defining the vehicle's behavior and managing the dynamics and responses based on operational inputs and environmental conditions.

Safety modules are designed to monitor the integrity and operational status of the vehicle, implementing fail-safes and checks that prevent malfunctions. Lastly, some modules do not neatly fit into these categories, encompassing functionalities that are unique or less common. Thus, our categories are defined as data transfer, navigation,  safety check, controller, and others. It is crucial to explicitly define each category for the LLM, recognizing that each model may have different understandings and interpretations of these categories, which influences their functionality. 

It is important to note that some modules may overlap across categories, such as those fitting both the navigation and controller categories, which manage route planning and vehicle control simultaneously. To evaluate the module retrieval performance of the LLM based on the provided ranking, we select the top k categories from the LLM's output, where k corresponds to the number of categories to which the module belongs.
To assess the effectiveness of the module retrieval unit, we employed precision, recall, and F1 score as our evaluation metrics. Table~\ref{tab:categorization_effectiveness} displays the performance results for each LLM. For each module category, we show the number of modules and the results of the evaluation metrics for each LLM. We calculate the precision, recall, and  $F_1$ score for module categorization based on summaries from both decompiled functions and normalized source code. The results from the normalized code serve as the upper bounds for each evaluation metric, as indicated in parentheses in Table~\ref{tab:categorization_effectiveness}. Our experimental results show that Codestral could outperform DeepSeek-Coder and CodeQwen.
Recall that, according to Table~\ref{tab:categorization_effectiveness}, the evaluation metrics for safety check modules are not high due to the limited number of these modules.


\subsection{\ref{enu:experiments:rq:time}: LLM Time Analysis}

Evaluating the time performance of the LLMs DeepSeek-Coder, CodeQwen, and CodeStral is crucial due to the significant time required for tasks such as summarizing decompiled functions and retrieving modules. As these LLMs are free to use, the main cost involved is the computational and operational time, which can be substantial. This is particularly important in environments where efficient resource utilization affects project timelines and costs. 

\begin{figure}[htbp]
    \centering
    \includegraphics[width=\columnwidth]{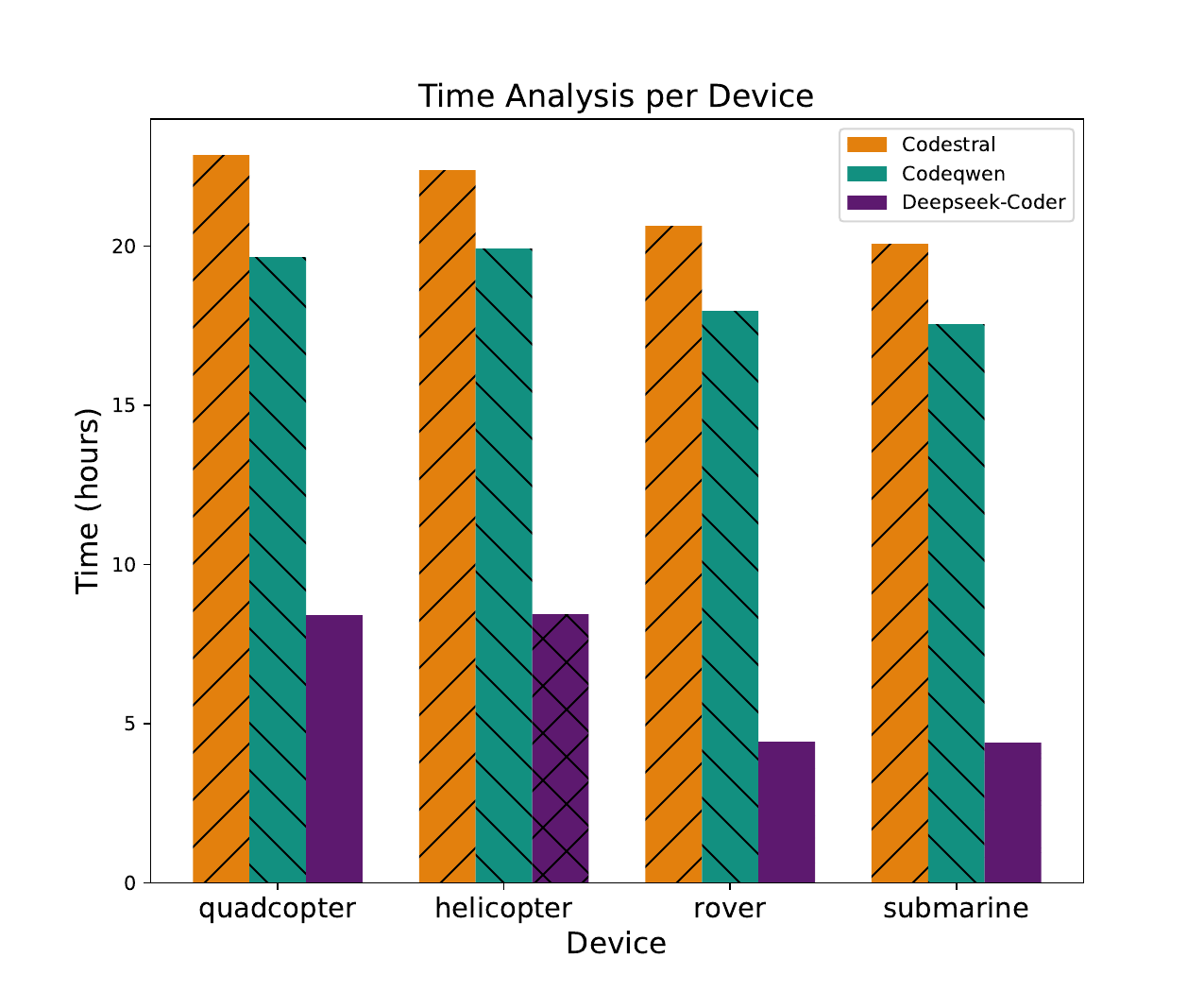}
    \caption{Comparison of time required for function summarization by three open-source LLMs (CodeStral, CodeQwen, and DeepSeek-Coder) across four ArduPilot devices. }
    \label{fig:timeanalysis}
\end{figure}

We calculate the total seconds each LLM requires to generate function summaries for all decompiled functions per device. We present this analysis in hours in Figure~\ref{fig:timeanalysis}. This evaluation indicates that CodeStral takes an average of 20 hours, while DeepSeek-Coder is the fastest, requiring approximately 5 hours to complete the summarization task.
We believe that the speed efficiency in DeepSeek-Coder
is because of our use of the structured (user vs. assistant) chat template style while prompting the LLM.
In our experience, this usually yields shorter outputs
from the language model. Note that this template was used to resolve some complications in DeepSeek-Coder in which inconsistencies in the special end token resulted in extremely long and repeated outputs.

The module categorization process is notably efficient, typically completed in a relatively short period and not more than two hours in the most extended cases. Given the minimal variability and the generally swift completion time, we have opted not to include a detailed analysis of time differences in this paper.

%% file: src/limitation.tex
\section{Challenges and Limitations}
\label{sec:limitations}
This section highlights the limitations of our approach.

\begin{itemize}
\item ArduPilot encompasses a comprehensive dataset with numerous functions and modules that often
  overlap across multiple categories. This overlap between module categories complicates our
  analysis because it can lead to inconsistencies in how modules are categorized.
\item Given the complexity and the specific nature of our dataset, we have deliberately limited the
  number of categories. Despite this limitation, the categories we have defined are highly useful
  and provide essential insights that researchers can leverage for effective analysis.
\item The categories we choose for our evaluation in this paper are tailored specifically to the
  domain of cyber-physical systems. Two objections might be raised in this context: First, that
  these categories are overly coarse-grained and that developers might want to see a more
  fine-grained categorization of functions within the binary. We chose these categories because they
  were of broad interest to CPS researchers, and also because of ease of establishing ground truth.
  The second objection is that they have a narrow focus on robotics firmware. This choice was purely
  a historical accident. However, we believe that the concept can be readily adapted to other
  domains. For example, malware analysts might be interested in categories such as network
  communication, key-logging, or process monitoring.
\item We also did not perform prompt engineering to optimize our results. It is possible that with
  more effort put into prompt design, one or more LLMs might achieve better accuracy.
\end{itemize}

%% file: src/future.tex
\section{Future Work}
\label{sec:future_work}
In this section, we outline our future plans for expanding both our evaluation and the conceptual framework of our research. 

\begin{itemize}
    \item We intend to broaden our evaluation by incorporating well-known and proprietary LLMs such as GPT-4, Copilot, and Claude into our assessment, comparing their effectiveness in function summarization and module retrieval.  
    \item As part of our future plans, we aim to enhance our evaluation by identifying the most effective prompts through a prompt engineering process. 
    \item We aim to enrich the semantic context of modules by using some tools such as SensorLoader~\cite{dasbach2023sensorloader}. These frameworks could potentially label modules with sensor-specific data flows or peripheral access patterns, enhancing the module categorization process.
    \item Our current system handles a limited number of general queries. Moving forward, we plan to refine our approach to accommodate more specialized queries, particularly those targeting specific functions within modules. For example, we aim to enable queries that identify well-known mathematical functions commonly utilized in cyber-physical systems. 
    \item We aim to further validate our approach by conducting evaluations on the PX4 dataset, widely recognized in the field of autopilot systems.
    
\end{itemize}

%% file: src/conclusion.tex
\section{Conclusion}

In this research, we design a code-based query system called ChatCPS to retrieve different module categories from stripped binary firmware.
ChatCPS first decomposes binary firmware into its constituent modules using the BCD algorithm. It then categorizes these modules based on function summaries from decompiled firmware functions, generated by three open-source LLMs: CodeQwen, CodeStral, and DeepSeek-Coder. This system can isolate specific functionalities within binary firmware, facilitating the analysis and understanding of its architecture. Our experimental results demonstrate that CodeStral surpasses other LLMs in module retrieval, achieving the best performance in our comparisons.